\documentclass[aps,prl,twocolumn,superscriptaddress,showpacs]{revtex4}
\usepackage{amsmath}
\usepackage{graphicx}
\usepackage{afterpage}
\usepackage{dcolumn}
\usepackage{bm}
\preprint{APS/123-QED}
\graphicspath{{graphs/}}
\begin{document}
\title{Optical properties of BaFe$_{2-x}$Co$_x$As$_2$}
\author{E. van Heumen}
\affiliation{D\'epartement de Physique de la Mati\`ere
Condens\'ee, Universit\'e de Gen\`eve, quai Ernest-Ansermet 24,
CH1211 , Gen\`eve 4, Switzerland}
\affiliation{Van der Waals-Zeeman Institute, University of
Amsterdam, NL-1018XE Amsterdam, The Netherlands}
\author{Y. Huang}
\author{S. de Jong}
\affiliation{Van der Waals-Zeeman Institute, University of
Amsterdam, NL-1018XE Amsterdam, The Netherlands}
\author{A.B. Kuzmenko}
\affiliation{D\'epartement de Physique de la Mati\`ere
Condens\'ee, Universit\'e de Gen\`eve, quai Ernest-Ansermet 24,
CH1211 , Gen\`eve 4, Switzerland}
\author{M.S. Golden}
\affiliation{Van der Waals-Zeeman Institute, University of
Amsterdam, NL-1018XE Amsterdam, The Netherlands}
\author{D. van der Marel}
\affiliation{D\'epartement de Physique de la Mati\`ere
Condens\'ee, Universit\'e de Gen\`eve, quai Ernest-Ansermet 24,
CH1211 , Gen\`eve 4, Switzerland}
\begin{abstract}
We present detailed temperature dependent optical data on BaFe$_{2-x}$Co$_{x}$As$_{2}$ (BCFA), with x = 0.14, between 4 meV and 6.5 eV. We analyze our spectra to determine the main optical parameters and show that in this material the interband conductivity already starts around 10 meV. We determine the superfluid density $\rho_{s}\approx2.2\pm0.5\cdot10^{7}$ cm$^{-2}$, which places optimally doped BFCA close to the Uemura line. Our experimental data shows clear signs of a superconducting gap with 2$\Delta_{1}$ = 6.2 $\pm$ 0.8 meV. In addition we show that the optical spectra are consistent with the presence of an additional band of strongly scattered carriers with a larger gap, 2$\Delta_{2}$ = 14 $\pm$ 2 meV.

\end{abstract}
\maketitle

Recently a new family of superconductors was discovered: the iron-pnictide superconductors \cite{kamihara-JACS-2008}. Several studies indicate that these materials are much closer to conventional Fermi liquids with moderate interaction strengths \cite{mazin-PRL-2008,lu-NAT-2008,chubukov-PRB-2008}, as compared to the strongly interacting, strongly correlated cuprate superconductors. The parent iron pnictides are metallic with the states at the Fermi level deriving from the Fe orbitals \cite{jong-PRB-2009} and, apart from an overall scaling factor of the bandwidth, appear to adhere to a conventional LDA framework \cite{lu-NAT-2008}. Superconductivity can be induced by chemical substitution or by applying hydrostatic pressure. The standard model of superconductivity explains the stability of the superconducting state as arising from an effective, attractive interaction between the electrons, resulting from overscreening of the Coulomb interaction by the electron-phonon interaction. However, if the superconducting order parameter has nodes (points in momentum space where the superconducting gap at the Fermi energy is zero and around which the phase of the order parameter changes sign), then the BCS gap equation allows solutions for repulsive interactions. In these cases the pairing can be mediated by electronic degrees of freedom or correlation effects. The determination of the order parameter symmetry is therefore an important clue to the nature of the interaction responsible for the pairing in novel superconductors. The symmetry of the superconducting gap in the pnictides is thought to be extended s-wave: momentum independent on each Fermi surface sheet, but with opposite phase on different sheets \cite{mazin-PRL-2008}. However, magnetic penetration depth studies find evidence for line nodes in some compounds \cite{fletcher-PRL-2009,hashimoto-arxiv-2009}. Several theoretical studies have suggested that line nodes can be formed in an otherwise uniform superconducting state by residual interactions between electron and hole bands \cite{chubukov-PRB-2009,parker-PRB-2009}. In angle resolved photoemission spectroscopy (ARPES) studies of K and Co doped BaFe$_{2}$As$_{2}$ (Ba122) momentum independent superconducting gaps are observed on several Fermi surface sheets \cite{evtushinsky-PRB-2009,terashima-PNAS-2009}. In this Letter we present optical data on BaFe$_{2-x}$Co$_{x}$As$_{2}$ (BFCA) with a critical temperature T$_{c}$ of 23 K. We observe the direct impact of superconductivity on the low frequency optical spectrum and estimate the superconducting gap to be $2\Delta \approx$ 6.2 $\pm$ 0.8 meV. We extract the superfluid density $\rho_{s}\approx2.2\pm0.5\cdot10^{7}$ cm$^{-2}$ which places BFCA close to the Uemura line. We also test whether our results support the presence of a second gap.

\begin{figure}[t!]
\centering
\includegraphics[width=8.5 cm]{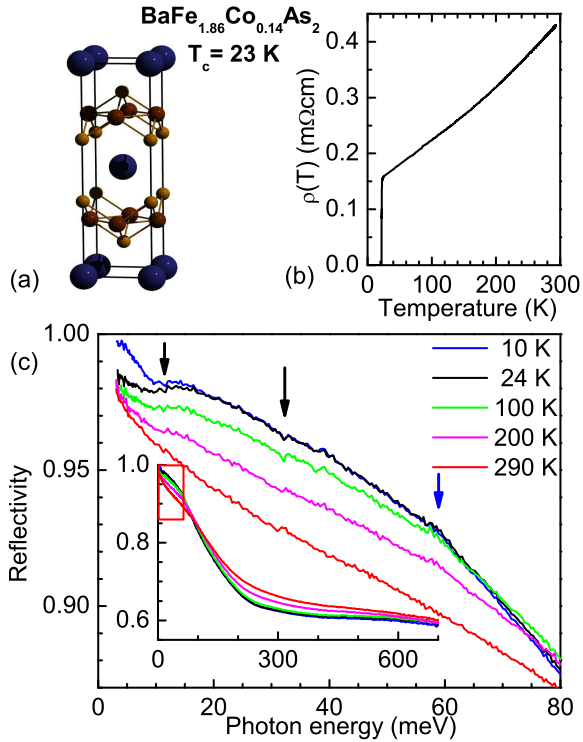}
\caption{(a): Unit cell of the tetragonal BaFe$_{2}$As$_{2}$ structure. (b): Temperature dependence of the resistivity. (c): Far infrared reflectance for selected temperatures. Note the strong increase in reflectance below 10 meV for data taken in the SC state (10 K). Black arrows indicate the two phonon absorption lines, while the blue arrow indicates a slope change at 60 meV. The inset shows the reflectance on a larger scale.}
\label{Refl}
\end{figure}

Single crystals of BFCA were grown from self-flux and the results presented here are obtained on an x = 0.14 Co doped sample. Co doping takes place in the Fe layers (see the crystal structure in fig. \ref{Refl}a). Our crystals exhibit a sharp superconducting (SC) transition at T$_{c}$ = 23 K as indicated by resistivity measurements (Fig. \ref{Refl}b). The resistivity shows no sign of a spin density wave transition at higher temperatures. The reflectivity has been measured between 4 meV and 0.75 eV. Ellipsometric measurements were made in the range 0.75 eV to 6.2 eV. Experiments were performed between 10 K and 300 K, using stabilized, high vacuum cryostats ($\le$ 10$^{-9}$ mbar in the MIR to UV range). We use a variational method to extract the optical conductivity in the entire range \cite{kuzmenko-RSI-2005}.

The reflectivity is shown in Fig. \ref{Refl}c. It scales at room temperature as $R(\omega)\approx 1-A\sqrt{\omega}$ for small frequencies, which is the Hagen-Rubens behavior expected for a metal. We observe two direct phonon absorption lines, indicated by black arrows. The lowest phonon mode, most likely related to vibrations of the Ba atoms \cite{reznik-arxiv-2008}, is observed at room temperature around 11 meV, but at lower temperatures it is masked by a stronger absorption arising from a low lying interband transition (see below). The second phonon we observe is most likely related to an Fe-As mode \cite{reznik-arxiv-2008}. It hardens from 31.5 meV at room temperature to 32.3 meV at 10 K.  With decreasing temperature several features appear in the far infrared reflectivity: (i) an absorption appears around 8 meV, (ii) a change in slope appears at 60 meV (indicated by the blue arrow in Fig. \ref{Refl}c) and (iii) below the critical temperature there is a clear impact of superconductivity on the reflectivity, pushing the reflectivity to unity below 5 meV. The reflectivity is shown on the entire measured range in the inset of Fig. \ref{Refl}c. 

\begin{figure}[t!]
\centering
\includegraphics[width=8.5 cm]{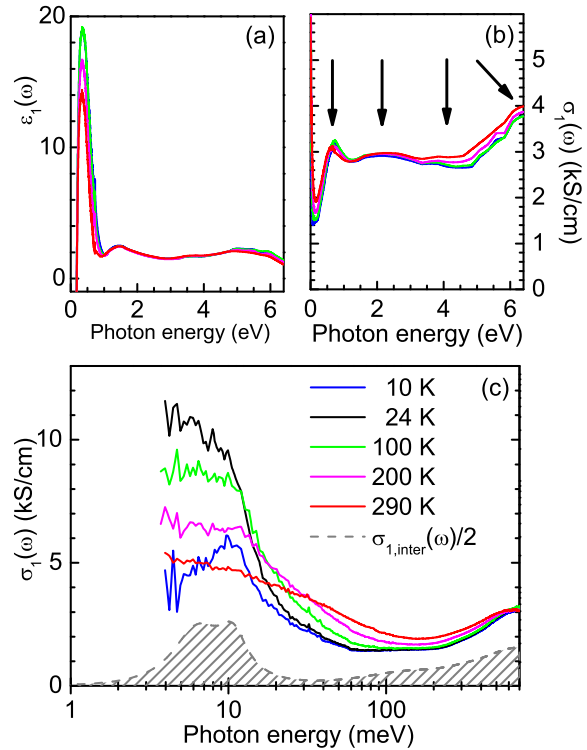}
\caption{(a): Dielectric function for selected temperatures (indicated in panel c). Above 750 meV the dielectric function is only weakly frequency dependent. (b): Optical conductivity at the same temperatures. We observe three interband transitions at 0.62 eV, 2.3 eV and 3.9 eV indicated by the arrows. A fourth interband transition falls just outside our experimental window. (c): Optical conductivity on a logarithmic energy scale. The shaded area indicates the contribution to the conductivity from low lying interband transitions at 24 K (divided by two for clarity of presentation).}
\label{epssig}
\end{figure}

The dielectric function, $\varepsilon_{1}(\omega)$, and optical conductivity, $\sigma_{1}(\omega)$, are shown in Fig. \ref{epssig}a and \ref{epssig}b. 
The optical conductivity is characterized by a narrow Drude peak and several interband transitions. We observe a broad structure around 2.3 eV, in agreement with previous work \cite{hu-PRL-2008} on the parent compound Ba122. The same peak is observed in LaFePO \cite{qazilbash-NATP-2009}, where it is seen to be much more pronounced. We observe another weak feature at 3.9 eV and the low energy tail of a transition which has a maximum above 6.2 eV, outside our experimental range. The low frequency conductivity is shown in Fig. \ref{epssig}c, more clearly displaying the Drude contribution. We observe several low energy absorptions which are more clearly visible at low temperatures, due to the narrowing of the Drude peak with decreasing temperature. The dashed gray area in Fig. \ref{epssig}c represents the interband contribution obtained from a Drude-Lorentz fit at 24 K (divided by two for clarity). The interband transitions in BFCA start around 10 meV with a strong absorption peak, followed by an onset around 100 meV and a distinct mid-infrared peak at 0.62 eV. The presence of these low lying transitions make it very difficult to separate the free charge carrier conductivity from the interband conductivity. A more detailed analysis of the interband conductivity is required for an analysis in terms of the extended Drude model and will be presented elsewhere. 

Having determined the main optical features we now turn our attention to the quantities that characterize the superconducting state. An important quantity measuring the robustness of the SC state is the superfluid density $\rho_{s}$. The presence of a superfluid density gives rise to a zero frequency condensate peak in the optical conductivity with an area $\rho_{s}=\omega^{2}_{p,s}$. Since the total spectral weight above and below the critical temperature has to remain constant, one can measure the superfluid density by analyzing the reduction of finite frequency spectral weight when the material becomes superconducting \cite{tinkham-PRL-1959}. From such an analysis we determine $\rho_{s}\approx2.2\pm0.5\cdot10^{7}$ cm$^{-2}$ (corresponding to a penetration depth of 340 nm). The same analysis shows that the superconductivity induced transfer of spectral weight is recovered around 40 meV. A second method to determine $\rho_{s}$ follows by making use of the Kramers-Kronig (KK) relations:  if a zero frequency $\delta$-peak of strength $\omega^{2}_{p,s}$ is present in $\sigma_{1}(\omega)$, applying the KK relations will lead to $\varepsilon_{1}(\omega)\propto -\omega^{2}_{p,s}/\omega^{2}$. Therefore the extrapolation of $-\omega^{2}\varepsilon_{1}(\omega)$ to zero frequency is a direct measure of the superfluid density. 
\begin{figure}[t!]
\centering
\includegraphics[width=8.5 cm]{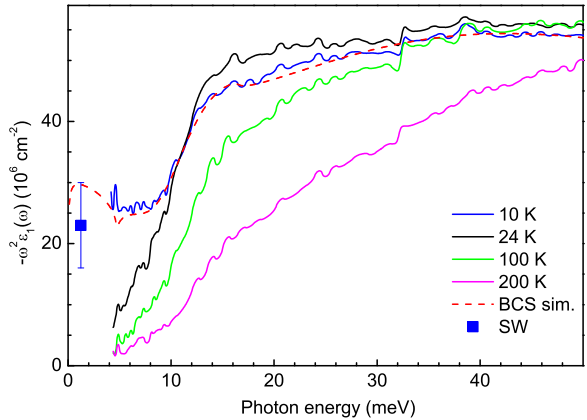}
\caption{$-\omega^{2}\varepsilon_{1}(\omega)$ for several temperatures. The extrapolation to zero frequency gives an estimate of the superfluid density. The blue square indicates $\rho_{s}$ estimated from a spectral weight analysis. The dashed line is the best fit result of the BCS analysis of the 10 K data. Both the experimental curve and the simulation show minimum at 2$\Delta$.}
\label{superfluid}
\end{figure}
This latter quantity is shown in Fig. \ref{superfluid}. We see that the extrapolation to zero of the experimental curve matches reasonably well with the value estimated from the spectral weight analysis. Estimates of the penetration depth $\lambda(0)\approx$ 208 nm on optimally doped BFCA crystals were reported in Ref. \cite{gordon-PRL-2009} based on measurements of H$_{c1}$. To compare this with our results one has to take into account the fact that our measurement is performed at 10 K. From the temperature dependence of $\Delta\lambda (T)$ also reported in \cite{gordon-PRL-2009} we obtain $\lambda(10 K)\approx$ 350 nm, in very good agreement with our observations. Therefore this system has a small superfluid density close to the Uemura line T$_{c}\propto\rho_{s}$, in sharp contrast with K doped Ba122, for which $\rho_{s}\approx22\cdot10^{7}$ cm$^{-2}$ with T$_{c}$ of only 36 K \cite{ren-PRL-2008}. 

The second quantity characterizing the superconducting state is the superconducting gap. At zero temperature the gap can be determined from the onset of absorption in the optical conductivity, but temperature broadening and the uncertainties related to the determination of the deviation of the reflectivity from unity make a reliable determination of the gap size difficult to realize. A method to determine the SC gap directly from experimental data was proposed by Marsiglio \textit{et al.} \cite{marsiglio-PRB-1996} based on the KK relations. They showed that BCS theory predicts a cusp like minimum in $-\omega^{2}\varepsilon_{1}(\omega)$ at $2\Delta$ for weak impurity scattering. Figure \ref{superfluid} shows such a minifmum around $2\Delta\approx$ 6.2 meV. Another conclusion following from the analysis in Ref. \cite{marsiglio-PRB-1996}, is that such a minimum occurs only if the order parameter has s-wave symmetry. The fact that this minimum is visible in our data therefore suggests s-wave symmetry. ARPES experiments have shown evidence for two gaps with different sizes on different Fermi surface sheets \cite{evtushinsky-PRB-2009,terashima-PNAS-2009}, but we observe no clear features that stem from a second gap.
\begin{figure}[t!]
\centering
\includegraphics[width=8.5 cm]{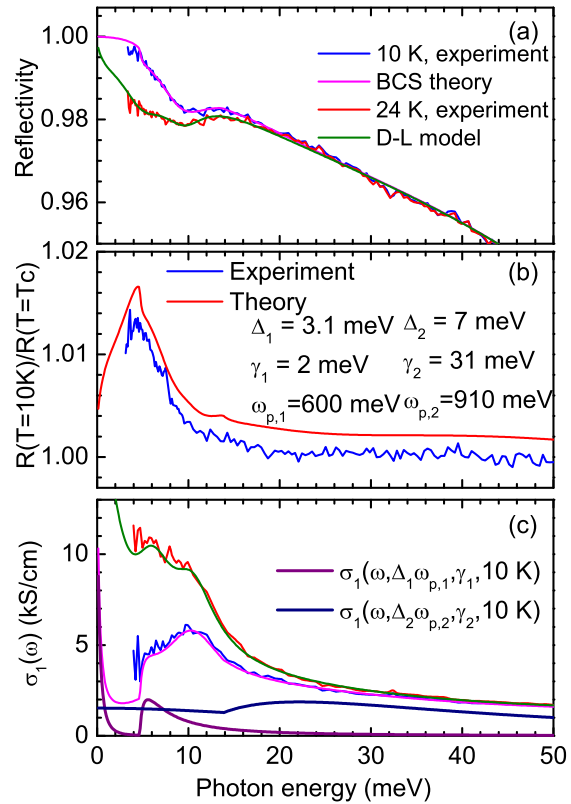}
\caption{(a): Low energy experimental reflectivity (blue) and reflectivity obtained from the  BCS simulation (violet) at 10 K. Also shown are the reflectivity (red) and Drude-Lorentz model result (green) at T$_{c}$. (b): Ratio of 10 K reflectivity to the one at T$_{c}$ (24 K) for the experimental data (blue) and simulation (offset for clarity by 1.002, in red). (c): Comparison of the 10 K optical conductivity determined from experiment (blue) and simulation (violet) and the 24 K data (red) and Drude - Lorentz model (green). Note the low frequency upturn in the simulation which arises from finite temperature broadening. We also show the two contributions to the conductivity from the BCS calculation.}
\label{BCS}
\end{figure}
To test whether our optical data supports the presence of two distinct gaps, we use an analysis based on a method put forward by Zimmermann \textit{et al.} \cite{zimmermann-physc-1991}. These authors derive expressions for the optical conductivity of a BCS superconductor with arbitrary impurity scattering, implemented in a fast computational routine that allows for direct least square optimization of the model parameters using experimental data (employing the standard Fresnel equations to obtain the reflectivity). To calculate the optical conductivity in this framework requires 4 input parameters: the reduced temperature $t = T/T_{c}$, gap value $\Delta$, plasma frequency $\omega_{p}$ and impurity scattering rate $\gamma=1/\tau$. Since we expect to have more than one band contributing to the conductivity, we assume that the conductivity can be described by the superposition of two independent channels: $\sigma(\omega,T)$ = $\sigma(\omega,\Delta_1,\gamma_1,\omega_{p,1},T)$ + $\sigma(\omega,\Delta_2,\gamma_2,\omega_{p,2},T)$. First we optimize the parameters of a standard Drude Lorentz model at 24 K, to obtain a good model for the interband optical conductivity. In the next step we take the gap value obtained above for the small gap and the larger one obtained from ARPES and STS experiments \cite{terashima-PNAS-2009,massee-PRB-2009} (i.e. $\Delta_1$ = 3.1 meV and $\Delta_2$ = 7 meV) and optimize the remaining parameters of the model. From this process we find $\omega_{p,1}$ = 0.6 eV, $\gamma_1$ = 2 meV, $\omega_{p,2}$ = 0.9 eV, and $\gamma_2$ = 31 meV. Fig. \ref{BCS}a shows the reflectivity data at 10 K and T$_{c}$ together with the result obtained from the method by Zimmermann \textit{et al.}. The parameters where optimized for the 10 K curve, while the result at T$_{c}$ was obtained from a Drude-Lorenz model. The gap induced features are more clearly seen in the ratio of the reflectivity at 10 K to that at T$_{c}$ (Fig \ref{BCS}b). The simulated curve has been offset for clarity. Figure \ref{BCS}c compares the optical conductivity with the BCS result. Note the substantial conductivity predicted by the simulation in the region below $2\Delta_{1}$, which arises from thermal excitations of unpaired electrons. By shifting the reflectivity up and down by 0.5 $\%$ and repeating the analysis, leaving the gap values as free parameters, we obtain error bars on the determined values: $\Delta_{1}$ = 3.1 $\pm$ 0.4 meV and $\Delta_{2}$ = 7 $\pm$ 1 meV. In Fig. \ref{BCS}c we also show the decomposition of the BCS conductivity in the individual conductivity channels. We see that the small gap contribution has a pronounced contribution to the conductivity, but the larger gap conductivity is rather featureless due to the large scattering rate. The assignment to the smaller gap at 2$\Delta_{1}$ = 6.2 meV is unambiguously supported by the optical spectra. These data are also consistent with the existence of a larger gap at 2$\Delta_{2}$ =14 meV, but this interpretation of the corresponding spectral features is probably not unique. The large gap obtained by ARPES experiments is found on a barrel around the $\Gamma$ point, which is connected by a nesting vector $Q=(\pi,\pi)$ (in the folded zone) to a barrel around $X$, exhibiting a smaller gap \cite{terashima-PNAS-2009}. In these experiments there is only one hole pocket around the $\Gamma$ point while there are two electron pockets around the $X$ point, which gives rise to a relative enhancement of the scattering rate on the hole pocket. This is in line with our observation of a larger scattering rate associated with the large gap. Interestingly, the absence of a clear optical signature of a second, larger gap was also the case in data on MgB$_{2}$ \cite{kuzmenko-PHYSC-2007}, where overwhelming evidence for the existence of two distinct gaps was found by other spectroscopies. 

In conclusion, we have presented a detailed analysis of the optical spectra of BFCA. This compound is characterized by a Drude peak and interband transitions starting around 10 meV. The superfluid density is much smaller compared to its hole doped cousin Ba$_{1-x}$K$_{x}$Fe$_{2}$As$_{2}$. We directly observe a small superconducting gap with 2$\Delta_{1}$ = 6.2 meV and our data are consistent with an additional gap at 2$\Delta_{2}$ = 14 meV. 

\acknowledgements
We would like to thank H. Luigjes, R. Huisman and A. de Visser for the resistivity measurement.
This work is supported by the Swiss National Science Foundation through Grant No. 200020-113293 and the National Center of
Competence in Research (NCCR) "Materials with Novel Electronic Properties", MaNEP, and is, in addition, part of the research programme of the Foundation for Fundamental Research on Matter (FOM), which is financially supported by the Netherlands Organisation for Scientific Research (NWO).

\bibliography{paper_citations}

\end{document}